\documentclass[twocolumn,showpacs,amsmath,amssymb,pra,aps]{revtex4}

\usepackage{amsmath}
\usepackage{graphicx}
\usepackage{bm}

\begin{document}

\title{
 Spontaneous decay of an excited atom placed near a rectangular plate
}

\author{Tuan Anh Nguyen}

\author{Ho Trung Dung}

\affiliation{
 Institute of Physics, Academy of Sciences and Technology,
 1 Mac Dinh Chi Street, District 1,
 Ho Chi Minh city, Vietnam
}

\date{Jan. 19, 2007}

\begin{abstract}
    Using the Born expansion of the Green tensor, we consider the
    spontaneous decay rate of an excited atom placed in the vicinity of
    a rectangular plate. We discuss the limitations of the commonly used
    simplifying assumption that the plate extends to infinity in the lateral
    directions and examine the effects of the atomic dipole moment
    orientation, atomic position, and plate boundary and thickness on the atomic
    decay rate. In particular, it is shown  that in the boundary region, the
    spontaneous decay rate can be strongly modified.
\end{abstract}


\maketitle


The ability to control the spontaneous decay process holds the key
to powerful applications in micro- and nano-optical devices.
Effective control can be achieved by tailoring the environment
surrounding the emitters. In theoretical analysis of surrounding
environment of different geometries, the most interesting ones
being of the resonator type, the boundary conditions are typically
taken into account only in directions in which the electromagnetic
field is confined or affected the most. For example, in a planar
configuration, only the boundary conditions in the normal
direction are taken into account while those in the lateral
directions are neglected (see, e.g., Ref. \cite{Chance78}). In a
cylindrical configuration that extends to infinity, the reverse is
true \cite{Erdogan93}. Under appropriate conditions, these
approximations are generally valid. However, as the sizes of
devices decrease and fall in the micro- and nano-meter ranges as
in the current trend of miniaturization, it is clearly of great
importance to keep track of the effects of all boundaries. One way
to calculate the spontaneous decay rate in an arbitrary geometry
is to directly solve the Maxwell equations using the finite
difference time domain method \cite{Xu00}. This method, which
relies entirely on numerical computation, is not without
weaknesses. It requires that the whole computational domain be
gridded, leading to very large computational domains in cases of
extended geometries, or in cases where the field values at some
distance are required. All curved surfaces must be modelled by a
stair-step approximation, which can introduce errors in the
results. Additionally, the discretization in time may be a source
of errors in the longitudinal field \cite{Xu00}.

Here we employ an approach that, in a sense, combines analytical
and numerical calculations, thereby significantly reducing the
numerical computation workload. This approach relies on first
writing the atomic decay rate in terms of the Green tensor
characterizing the surrounding media \cite{Agarwal75,Ho00}.
Although this formula holds for arbitrary boundary conditions,
exact analytical evaluation of the Green tensors for realistic,
finite-size systems can be very cumbersome or even prohibitive.
Following \cite{Buhmann06} and \cite{Ho06}, where the atom-body
van der Waals force and the local-field correction, respectively,
have been considered, we circumvent the task of an exact
calculation of the Green tensor by writing it in terms of a Born
series and restrict ourselves to leading-order terms. The boundary
conditions enter the theory only via the integral limits. This
approach is universal in the sense that it works for an arbitrary
geometry of the surrounding media, and can be used to evaluate any
characteristics of the matter-electromagnetic field interaction
expressible in terms of the Green tensor. In this paper, we are
concerned mostly with the spontaneous decay rate of an excited
atom placed near a rectangular plate. Our aim is twofold: first,
we compare our results with those for an infinitely extended plate
in order to establish in a quantitative way the conditions under
which the approximation of an infinitely extended plate is valid;
second, we examine the effects brought about by the presence of
the boundaries in the lateral directions.


The (classical) Green tensor of an arbitrary dispersing and
absorbing body satisfies the equation
\begin{align}
\label{e1}
     &\hat{H} {\bm G}({\bf r},{\bf r}',\omega)
     = \delta( {\bf r}-{\bf r}') \bm{I},
\\
\label{e2}
     &\hat{H}({\bf r}) \equiv
     \bm{\nabla}\times \frac{1}{\mu({\bf r},\omega)} \bm{\nabla}\times
     -\frac{\omega^2}{c^2}\,\varepsilon({\bf r},\omega),
\end{align}
($\bm{I}$-unit tensor) together with the boundary condition at
infinity, where $\varepsilon({\bf r},\omega)$ [$\mu({\bf
r},\omega)$] is the frequency- and space-dependent complex
permittivity (permeability) which obeys the Kramers-Kronig
relations.

Decomposing the permittivity and permeability as
\begin{equation}
\label{e3}
     \varepsilon({\bf r},\omega)
     =\bar{\varepsilon}({\bf r},\omega)
     +\chi_\varepsilon({\bf r},\omega) ,
     \ \
     \mu({\bf r},\omega)
     =\bar{\mu}({\bf r},\omega)
     +\chi_\mu({\bf r},\omega) ,
\end{equation}
and assuming that the solution $\bar{\bm{G}}({\bf r},{\bf
r}',\omega)$ to the equation
$
     \hat{\bar{H}}({\bf r}) \bar{\bm G}({\bf r},{\bf r}',\omega)
     = \delta( {\bf r}-{\bf r}') \bm{I},
$
where $\hat{\bar{H}}$ is defined as in Eq. (\ref{e2}) with
$\bar{\varepsilon}$ instead of $\varepsilon$ and $\bar{\mu}$
instead of $\mu$, is known, the Green tensor $\bm{G}$ can be
written as
\begin{equation}
\label{e6}
    \bm{G}(\mathbf{r},\mathbf{r}',\omega)
    =\bar{\bm{G}}(\mathbf{r},\mathbf{r}',\omega)
    +\bm{G}'(\mathbf{r},\mathbf{r}',\omega).
\end{equation}
Substituting Eq. (\ref{e6}) into Eq. (\ref{e1}) and using the
identity $(\bar{\mu}+\chi_\mu)^{-1}=\bar{\mu}^{-1}
\sum_{l=0}^\infty (\chi_\mu/\bar{\mu})^l$, it can be found
that
\begin{align}
\label{e7}
    &\hat{H}({\bf r}) \bm{G}'(\mathbf{r},\mathbf{r}',\omega)
    = \hat{H}_\chi({\bf r})
    \bar{\bm{G}}(\mathbf{r},\mathbf{r}',\omega) \equiv
    \tilde{\bm{G}}(\mathbf{r},\mathbf{r}',\omega)
\\
\label{e7a}
    &\hat{H}_\chi({\bf r}) \equiv
     - \bm{\nabla}\times \frac{1}{\bar{\mu}({\bf r},\omega)}
     \sum_{l=1}^\infty
     \frac{\chi^l_\mu({\bf r},\omega)}{\bar{\mu}^l({\bf r},\omega)}
     \bm{\nabla}\times
     +\frac{\omega^2}{c^2}\chi_\varepsilon({\bf r},\omega),
\end{align}
i.e., $\bm{G}'$ satisfies the same differential equation as the
one governing the electric field, with the current being equal to
$\tilde{\bm{G}}$. Hence it can be written as a convolution of this
current with the kernel $\bm{G}$:
$    \mathbf{G'}(\mathbf{r},\mathbf{r}',\omega)
    =\int d^3s \bm{G}(\mathbf{r},\mathbf{s},\omega)
    \tilde{\bm{G}}(\mathbf{s},\mathbf{r}',\omega) \,.
$
Substitution of $\bm{G}'$ in this form in Eq. (\ref{e6}) and iteration
lead to the desired Born series
\begin{align}
\label{e10}
     &\bm{G}({\bf r},{\bf r}',\omega) =
      \bar{\bm{G}}({\bf r},{\bf r}',\omega)
      + \sum_{k=1}^\infty \bm{G}_k({\bf r},{\bf r}',\omega),
\\
\label{e11}
     &\bm{G}_k({\bf r},{\bf r}',\omega) =
      \Biggl(\prod_{j=1}^k\int\mathrm{d}^3s_j \Biggr)
\nonumber\\
     &\quad\times
      \bar{\bm{G}}({\bf r},{\bf s}_1,\omega)
      \tilde{\bm{G}}({\bf s}_1,{\bf s}_2,\omega) \cdots
      \tilde{\bm{G}}({\bf s}_k,{\bf r}',\omega) \, .
\end{align}

This formal expansion for the Green tensor is valid for an
arbitrary geometry, permittivity, and permeability of the macroscopic bodies.
Obviously, one of the situations in which the Born series is
particularly useful is when $\chi_\varepsilon$ is a perturbation
to $\bar{\varepsilon}$ and $\chi_\mu$ is a perturbation to
$\bar{\mu}$, thereby one makes only a small error cutting off
higher-order terms. For such weakly dielectric and magnetic
bodies, it is natural to choose
\begin{equation}
\label{e12}
      \bar{\varepsilon}({\bf r},\omega)
      = \bar{\mu}({\bf r},\omega)=1
\end{equation}
with \mbox{$\chi_\lambda({\bf r},\omega)$ $\!=$ $\!\chi_{\lambda
{\rm R}}({\bf r},\omega)+i\chi_{\lambda{\rm I}}({\bf r},\omega)$},
$|\chi_\lambda({\bf r},\omega)|\ll 1$ $(\lambda=\varepsilon,\
\mu)$ [cf. Eqs. (\ref{e3})]. This implies we restrict ourselves to
frequencies far from a medium resonance.

The Green tensor $\bar{\bm G}$ corresponding to
$\bar{\varepsilon}({\bf r},\omega)=\bar{\mu}({\bf r},\omega)=1$ is
the vacuum Green tensor
\begin{align}
\label{e13}
     &\bar{\bm{G}}({\bf r},{\bf r}',\omega)
     = - \frac{\delta({\bf u}) }
        {3 k^2} \bm{I}
     + \frac{k}{4\pi}
       (a\bm{I} - b \hat{{\bf u}} \otimes \hat{{\bf u}})
       e^{iq},
\\[1ex]
\label{e14}
   &a \equiv a(q) = \frac{1}{q} + \frac{i}{q^2} - \frac{1}{q^3},
\quad
   b \equiv b(q) = \frac{1}{q} + \frac{3i}{q^2} - \frac{3}{q^3}
\end{align}
 ($k$ $\!=$ $\omega/c$; ${\bf u}$ $\!\equiv$ $\!{\bf r}-{\bf r}'$;
$\hat{{\bf u}}={\bf u}/u$, and $q$ $\!\equiv$ $ku$).

A description of quantities such as the emission pattern or the
interatomic van der Waals forces requires the knowledge of the
Green tensor of different positions, while a description of
quantities such as the spontaneous decay rate of an excited
atom or the atom-body van der Waals forces requires the knowledge of
the Green tensor of equal positions. Substituting Eq. (\ref{e13})
in Eq. (\ref{e11}) and assuming that the position ${\bf r}$
lies outside the region occupied by the macroscopic bodies, we
derive for the first-order term in the Born expansion of the
equal-position Green tensor
\begin{align}
\label{e17}
     &\bm{G}_1({\bf r},{\bf r},\omega) =
      \frac{k^2}{16\pi^2}
      \int\mathrm{d}^3s\,\hat{H}_{\chi 1}({\bf s})
\nonumber\\
     &\quad\times
       [a^2\bm{I} + (b^2-2ab)
            \hat{{\bf u}} \otimes \hat{{\bf u}}]
       e^{2iq}
\end{align}
[${\bf u}$ $\!\equiv$ $\!{\bf r}-{\bf s}$; $q=ku$; $a$ $\!=$
$\!a(q)$; $b$ $\!=$ $\!b(q)$;
$\hat{H}_{\chi 1}({\bf r}) \equiv
 - \bm{\nabla}\times \chi_\mu({\bf r},\omega)
 \bm{\nabla}\times
 +\frac{\omega^2}{c^2}\chi_\varepsilon({\bf r},\omega)$ --
linear part of $\hat{H}_\chi$, Eq. (\ref{e7a})].
Higher-order terms can easily be
obtained by repeatedly using Eq. (\ref{e13})  in Eq. (\ref{e11}).


\begin{figure}[!t!]
\noindent
\includegraphics[width=0.7\linewidth]{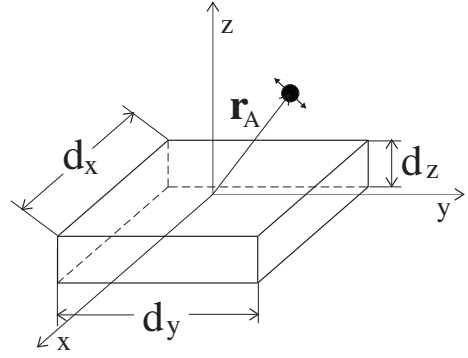}
\caption{
  A dipole emitter in the vicinity of a rectangular plate.
  }
\label{scheme}
\end{figure}%
%

Our system consists of an excited two-level atom surrounded by
macroscopic media, which can be absorbing and dispersing. In the
electric-dipole and rotating-wave approximations, the atomic decay
rate reads as \cite{Agarwal75,Ho00}
\begin{equation}
\label{e18}
     \Gamma = \frac{2k_{\rm A}^2}{ \hbar\varepsilon_0}\,
     {\bf d}_{\rm A} {\rm Im}\,{\bm G}
     ({\bf r}_{\rm A},{\bf r}_{\rm A},\omega_{\rm A})
     {\bf d}_{\rm A},
\end{equation}
where ${\bf d}_{\rm A}$ and $\omega_{\rm A}$ are the atomic dipole
and shifted transition frequency, $k_{\rm A}=\omega_{\rm A}/c$,
and ${\bm G}$ is the Green tensor describing the surrounding
media.

%
\begin{figure}[!t!]
\noindent
\includegraphics[width=0.7\linewidth]{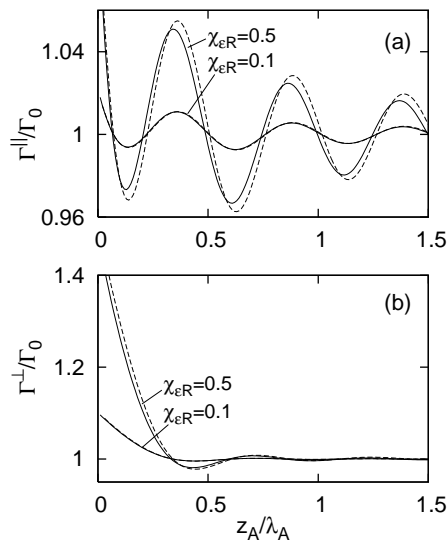}
\caption{
  Atom-surface distance dependence of the normalized spontaneous decay rate
  of an excited atom positioned at ($0,0,z_{\rm A}$) near an infinite planar plate (solid
  line) and a rectangular plate ($d_x=d_y=10\lambda_{\rm A}$, dashed line) of equal thickness
  $d_z=0.2\lambda_{\rm A}$ and equal $\chi_\varepsilon=\chi_{\varepsilon{\rm R}}+i10^{-8}$.
  Case (a) is for a $x$-oriented dipole moment,
  while case (b) is for a $z$-oriented dipole moment.
  }
\label{chi}
\end{figure}%
%
In accordance with the linear Born expansion, Eqs. (\ref{e10}), (\ref{e13}), and
(\ref{e17}) yield, for a purely electric material,
\begin{align}
\label{e20}
     &\frac{\Gamma^{\parallel(\bot)}}{\Gamma_0} = 1
      +\frac{3k_{\rm A}^3}{8\pi} {\rm Im} \Biggl\{
      \int\mathrm{d}^3s \,\chi_\varepsilon({\bf s},\omega_{\rm A})
\nonumber\\
      &\times
       \biggl[ a^2 + (b^2-2ab)\frac{1}{u^2}
       {(x-x_{\rm A})^2\atop (z-z_{\rm A})^2}\biggr]
       e^{2iq} \Biggr\}
\end{align}
 [$\Gamma_0=k_\mathrm{A}^3d_\mathrm{A}^2/
 (3\pi\hbar\varepsilon_0)$ - free-space decay rate, ${\bf s}=(x,y,z)$, $u=|{\bf
s}-{\bf r}_{\rm A}|$, $q=k_{\rm A}u$, $a=a(q)$, $b=b(q)$] for
\mbox{$x$-($z$-)}oriented dipole moments. Equations (\ref{e20})
are our main working equations. Just like the Born expansion of
the Green tensor, they hold for an arbitrary geometry of the
surrounding environment.

Next let us be specific about the shape of the macroscopic bodies.
We consider a rectangular plate of dimensions $d_x$, $d_y$, and
$d_z$ and choose a Cartesian coordinate system such that its
origin is located at the center of one surface of the plate, as
sketched in Fig. \ref{scheme}. Then $\Gamma^\parallel$ represents
the spontaneous decay rate of a dipole moment parallel to a plate
surface and $\Gamma^\bot$ -- that of a dipole moment normal to the
same surface. Since no further analytical calculation in Eqs.
(\ref{e20}) seems possible, we resort to numerical computation.

In Fig. \ref{chi} we present the spontaneous decay rate in
accordance with the linear Born expansion (\ref{e20}), as a
function of the atom-surface distance for two different values of
the permittivity. The same quantity but for an atom placed near an
infinitely extended planar slab is plotted using the Green tensor
given in Ref. \cite{Tomas95}. It can be seen that when the lateral
dimensions of the rectangular plate are sufficiently large and the
absolute value of the permittivity is sufficiently close to one
(the case of $\chi_\varepsilon=0.1+i10^{-8}$ in the figure), the two results
almost coincide for both dipole moment orientations. As $\chi_{\varepsilon{\rm R}}$
increases, the agreement worsens but is still quite good at
$\chi_{\varepsilon{\rm R}}=0.5$. Further numerical calculations indicate that
in the range of $|\chi_\varepsilon(\omega_{\rm A})| \stackrel{<}{\sim} 0.5$, the
spontaneous decay rate can be well approximated by the linear Born
expansion. In Fig. \ref{chi}, the atom has been moved along the
$z$-axis with $x_{\rm A}=y_{\rm A}=0$. When the atom is moved along
other lines parallel to the $z$-axis but nearer to the border, the
$z_{\rm A}$-dependence of the normalized spontaneous decay rates behaves
in a similar way as in Fig. \ref{chi} but with its value being
generally closer to one. Having determined the range of
permittivities where the linear Born expansion provides a good
approximation to the spontaneous decay rate, we move on now to
investigate when the plate can be regarded as an infinite slab.

\begin{figure}[!t!]
\noindent
\includegraphics[width=0.7\linewidth]{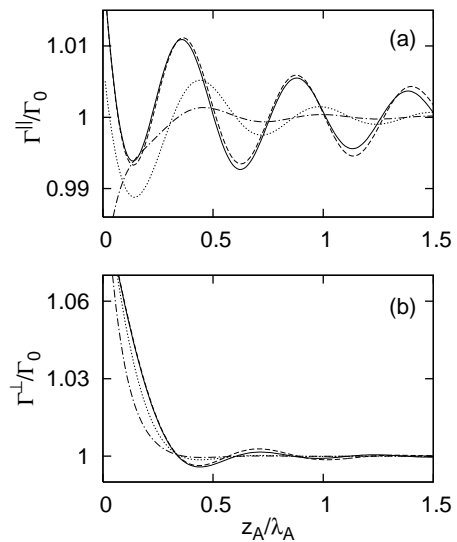}
\caption{
  The same as in Fig. \ref{chi} but for different sizes of the
  rectangular plate: $d_x=d_y=3\lambda_{\rm A}$
  (dashed line), $0.4\lambda_{\rm A}$ (dotted line), and
  $0.2\lambda_{\rm A}$ (dash-dotted line). In the last case,
  the plate is actually a cube. The curves for an infinite planar
  plate are shown by solid line and $\chi_\varepsilon=0.1+i10^{-8}$.
  }
\label{ver}
\end{figure}%
%

\begin{figure}[!t!]
\noindent
\includegraphics[width=0.7\linewidth]{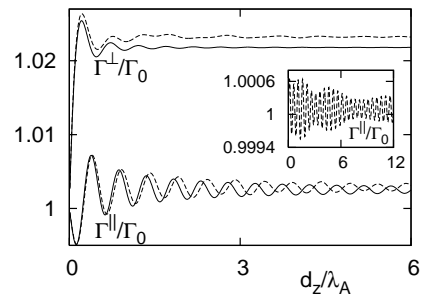}
\caption{
  Plate-thickness dependence of the spontaneous decay rate
  of an excited atom located near an infinitely extended planar plate (solid
  line) and a rectangular plate ($d_x=d_y=10\lambda_{\rm A}$, dashed line) of equal
  $\chi_\varepsilon=0.1+i10^{-8}$. The atomic position is fixed at (0,0,$0.2\lambda_{\rm A}$)
  in the main figure, and (0,0,$5\lambda_{\rm A}$) in the inset.
  }
\label{thick}
\end{figure}%
%

In Fig. \ref{ver}, we gradually reduce the lateral sizes of the
rectangular plate while keeping its thickness constant, and
compare the resulting spontaneous decay rates with that for an
infinite slab. To be on the conservative side, the permittivity is
set equal to $\varepsilon(\omega_{\rm A})=1.1+i10^{-8}$ -- a value which
is very close to one (cf. Fig. \ref{chi}). For lengths of the
lateral sides comparable to the atomic transition wavelength, the
infinite slab approximation starts to differ noticeably from the
linear Born expansion (see the figure, case of
$d_x=d_y=3\lambda_{\rm A}$), which in this situation is regarded
as a good and nondegrading approximation.  As the lateral sizes of
the plate decrease further and become smaller than $\lambda_{\rm A}$,
the infinite-slab approximation fails completely (see cases of
$d_x=d_y=0.4\lambda_{\rm A}$ and $0.2\lambda_{\rm A}$ in the figure). In
other words, while a rectangular plate with lateral sizes much larger
than the atomic transition wavelength can be more or less
treated as an infinite slab, care should be taken when the sizes
are reduced to about or below a wavelength. This happens for both
normal and parallel to the surface dipole moment orientations.
When the rectangular plate can be roughly regarded as an infinite
slab, the infinite-slab curve and the linear Born expansion curve
agree better when the atom is placed closer to the surface (see
dashed and solid curves in the figure). This can be explained by
that the closer to the surface the atom is situated, the more it
is inclined to see the plate as infinite.

%
\begin{figure}[!t!]
\noindent
\includegraphics[width=0.7\linewidth]{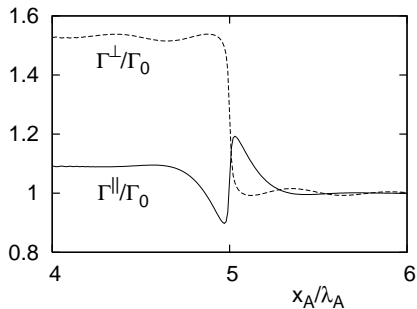}
\caption{
  Effects of the presence of a boundary in the $x$-direction on the
  spontaneous decay rate of an excited atom located near
  a rectangular plate of permittivity
  $\varepsilon(\omega_{\rm A})=1.5+i10^{-8}$ and dimensions
  $d_z=0.2\lambda_{\rm A}$ and $d_x=d_y=10\lambda_{\rm A}$. The atom is located
  at ($x_{\rm A}$,0, $0.01\lambda_{\rm A}$).
  }
\label{hor}
\end{figure}%
%

Besides the dependence on the lateral sizes, whether a rectangular
plate can be treated as an infinitely extended one clearly depends
on its thickness as well. It is intuitively obvious that even when
plate lateral sizes are much larger than the atomic transition
wavelength, the plate cannot be treated as extending to infinity
if its thickness is comparable with the lateral sizes. In Fig.
\ref{thick}, we compare the $d_z$-dependence of the spontaneous
decay rate for a rectangular plate with that for an infinite slab.
The agreement between the two curves, being very good for
sufficiently thin plates, gradually worsens with an increasing
plate thickness. The disagreement is already noticeable at
$d_z\sim \lambda_{\rm A}$ -- a value which is still much smaller than
the lateral sizes $d_x=d_y=10\lambda_{\rm A}$, and it sets in earlier
for $\Gamma^\bot$ than for $\Gamma^\parallel$. The two
calculations predict quite different large-thickness limits. This
means that in the case of thick plates, one must take into account
the boundary conditions in the lateral directions in order to
obtain reliable results.

Account of the boundary conditions in the
lateral directions also gives rise to some curious beating in the
$d_z$-dependence of the spontaneous decay rate, which is
especially visible when the atom has a dipole moment oriented
parallel to the surface and is situated somewhat away from the
surface (see Fig. \ref{thick}, inset). Let's have a closer look at,
say, the decay rate for a dipole
oriented parallel to the surface.  Using the stationary phase method,
one obtains from Eq. (\ref{e20})
\begin{align}
\label{e21}
     &\frac{\Gamma^\parallel}{\Gamma_0} \simeq 1
      +\frac{3k_{\rm A}^3}{2\pi} {\rm Im} \Biggl[
      \chi_\varepsilon \int_0^{d_z}{\rm d} z a^2(q_z) e^{2iq_z}
\nonumber\\
      &\times
      \int_0^{\frac{d_x}{2}} {\rm d}x \int_0^{\frac{d_y}{2}} {\rm d}y \,
      e^{i\frac{k_{\rm A}^2}{q_z}(x^2+y^2)}
        \Biggr]
\end{align}
[$q_z=k_{\rm A}(z+z_{\rm A})$]. The integrals over $x$ and $y$ in
Eq. (\ref{e21}) contain Fresnel integrals. In the limit of an infinite plate
$d_x,\ d_y \rightarrow \infty$, Eq. (\ref{e21}) becomes
\begin{align}
\label{e22}
     &\frac{\Gamma^\parallel}{\Gamma_0} \simeq 1
      +\frac{3k_{\rm A}}{16} {\rm Im} \Biggl[
      \chi_\varepsilon \int_0^{d_z}{\rm d} z a^2(q_z)q_z e^{2iq_z} (1+i)^2
      \Biggr].
\end{align}
As a function of $z/\lambda_{\rm A}$, the integrand
in Eq. (\ref{e22}) has a period of $\frac{1}{2}$.
Since this period is $z$-independent, the resulting integral must exhibit oscillations with the same
period, as confirmed by Fig. \ref{thick}, solid curves.
These oscillations survive for plates of finite lateral extensions (Fig. \ref{thick}, dashed curves).
The beating clearly arises from the finite values of $d_x$ and $d_y$.
Note that as a function of $z$, the inner integrands in Eq. (\ref{e21})
have a `period' that is $z$- dependent and that increases with increasing $z$.

Next we turn to elucidating the influence of the boundaries in the
$x$- and $y$-directions on the spontaneous decay rates. As can be
seen from Fig. \ref{hor}, where an edge is present at
$x_{\rm A}=5\lambda_{\rm A}$, the decay rates exhibit oscillations near the
boundary with a particularly strong magnitude right on either side
of it, and damping tails. The oscillations are more pronounced for
a dipole moment oriented parallel to the ($xy$)-plane than for a
$z$-oriented dipole moment. One can notice that when the
projection of the atomic position on the $xy$-plane lies outside
and sufficiently far from the boundaries, the spontaneous decay
rate approaches one in free space, as it should.


In summary, using the Born expansion of the Green tensor, we have
considered the decay rate of an atom located near a plate of
rectangular shape. We have shown that a rectangular plate can be
treated as extending to infinity only when its lateral sizes are
much larger than the atomic transition wavelength, its thickness
sufficiently small, and the atom is located close enough to the
plate surface.  We have also shown that a boundary in the lateral
directions can give rise to significant modifications of the decay
rate in either side of it. The first-order Born expansion remains
quite reliable even at a value of permittivity as high as 1.5.
Inclusion of higher-order terms would allow one to investigate
more dense media.

H.T.D. thanks S. Y. Buhmann and D.-G. Welsch for enlightening discussions.
We are grateful to J. Weiner for a critical reading of the manuscript.
This work has been supported by the Ho Chi Minh city National
University and the National Program for Basic Research of Vietnam.

\end{document}